\documentclass{article}
\usepackage{spconf,amsmath,graphicx,hyperref}
\usepackage[utf8]{inputenc} 
\usepackage[T1]{fontenc}
\usepackage{enumitem}
\setlist{nosep, leftmargin=14pt}
\usepackage{mwe} 
\usepackage{hyperref}       
\usepackage{url}            
\usepackage{booktabs}       
\usepackage{amsfonts}       
\usepackage{nicefrac}       
\usepackage{microtype}      
\usepackage{lipsum}
\usepackage{stackengine}
\usepackage{float}
\usepackage{amssymb}
\usepackage{subfig}
\usepackage{float}
\usepackage{color, colortbl}
\usepackage{rotating}
\usepackage{subfig}
\usepackage{tikz}
\usepackage{xcolor}
\usepackage{animate}
\usepackage{mathrsfs}
\usepackage{multirow}

\usepackage[ruled,vlined]{algorithm2e}

\hypersetup{
    colorlinks,
    linkcolor={red!50!black},
    citecolor={blue!80!black},
    urlcolor={blue!80!black}
}

\title{Efficient Brain Network Estimation with Sparse ICA in Non-Human Primate Neuroimaging}

\name{Qiang Li$^{1}$, Liang Ma$^{1}$, Masoud Seraji$^{1,2}$, Shujian Yu$^{3}$, Yun Wang$^{4,5}$, Jingyu Liu$^{1,6}$, Vince D. Calhoun$^{1,6}$}
\address{$^{1}$Tri-Institutional Center for Translational Research in Neuroimaging and Data Science (TReNDS), \\
Georgia State, Georgia Tech, and Emory University, Atlanta, GA, United States \\
$^{2}$School of Psychology, University of Texas at Austin, Austin, TX, United States\\
$^{3}$Department of Computer Science, Vrije Universiteit Amsterdam, The Netherlands\\
$^{4}$Department of Computer Science, Emory University, Atlanta, GA, United States\\
$^{5}$Department of Biomedical Informatics, Emory University, Atlanta, GA, United States\\
$^{6}$Department of Computer Science, Georgia State University, Atlanta, GA, United States} 
\begin{document}
\maketitle

\begin{abstract}
Independent component analysis (ICA) is widely used to separate mixed signals and recover statistically independent components. However, in non-human primate neuroimaging studies, most ICA-recovered spatial maps are often dense. To extract the most relevant brain activation patterns, post-hoc thresholding is typically applied-though this approach is often imprecise and arbitrary. To address this limitation, we employed the Sparse ICA method, which enforces both sparsity and statistical independence, allowing it to extract the most relevant activation maps without requiring additional post-processing. Simulation experiments demonstrate that Sparse ICA performs competitively against 11 classical linear ICA methods. We further applied Sparse ICA to real non-human primate neuroimaging data, identifying several independent component networks spanning different brain networks. These spatial maps revealed clearly defined activation areas, providing further evidence that Sparse ICA is effective and reliable in practical applications.
\end{abstract}


\section{Introduction}
\label{sec:intro}
\subsection{Independent Component Analysis}
Independent component analysis (ICA) is a well-known blind source separation (BSS) technique used to decompose multivariate signals into latent components that are statistically independent~\cite{hyvarinen1999fast,bell1995information,Hyvrinen2013IndependentCA}. Given observed data \( \mathbf{X} \in \mathbb{R}^{n \times p} \), where each row represents an observation and each column a measured variable, ICA assumes a linear generative model:

\begin{equation}
\mathbf{X} = \mathbf{A} \mathbf{S}    
\end{equation}

where \( \mathbf{A} \in \mathbb{R}^{n \times q} \) is an unknown mixing matrix, and \( \mathbf{S} \in \mathbb{R}^{q \times p} \) contains the latent source signals assumed to be mutually independent. The goal of ICA is to estimate a demixing matrix \( \mathbf{W} \in \mathbb{R}^{q \times n} \) such that:

\begin{equation}
\hat{\mathbf{S}} = \mathbf{W} \mathbf{X}    
\end{equation}

Statistical independence is often enforced by maximizing non-Gaussianity (e.g., kurtosis or negentropy)~\cite{hyvarinen1999fast} or by minimizing multivariate mutual information~\cite{bell1995information}, also known as total correlation (TC)~\cite{watanabe1960information}. In practice, the data is typically preprocessed through centering and whitening, ensuring \( \mathbb{E}[\mathbf{X}] = \mathbf{0} \) and \( \operatorname{Cov}(\mathbf{X}) = \mathbf{I} \), which simplifies the estimation process.

\subsection{Sparsity versus Statistical Independence}
Classical ICA seeks to recover latent components that are statistically independent, typically in the sense of non-Gaussianity~\cite{Hyvrinen2013IndependentCA}. However, the spatial maps obtained via classic ICA are often dense, meaning that many of their entries are small but nonzero. To emphasize only the most relevant activations, it is common to apply post-hoc thresholding to these dense maps.

Sparse ICA, by contrast, introduces a hard sparsity directly into the estimation process. This is often achieved by augmenting the ICA objective function with an \(\ell_1\)-norm penalty on the components \( \hat{\mathbf{S}} \), encouraging many entries to be exactly zero. The resulting optimization problem takes the form:

\begin{equation}    
\min_{\mathbf{W}} \; \mathcal{L}_{\text{ICA}}(\mathbf{W} \mathbf{X}) + \lambda \| \mathbf{W} \mathbf{X} \|_1
\end{equation}

where \( \mathcal{L}_{\text{ICA}} \) measures statistical dependence (e.g., via mutual information or negentropy), and \( \lambda > 0 \) controls the level of sparsity. This approach yields components that are both independent and sparse, enhancing interpretability and reducing the need for heuristic post-processing.

\subsection{Our Contributions}
In this study, we incorporate Sparse ICA, designed to jointly enforce sparsity, and also statistical independence in component estimation, while maintaining computational efficiency, to non-human primate neuroimaging. This method was originally proposed in the previous work~\cite{wang2024sparse}, where it was evaluated against widely used ICA algorithms including FastICA~\cite{hyvarinen1999fast} and Infomax ICA~\cite{bell1995information}. In this extended study, we broaden our evaluation to include 11 additional classical ICA algorithms, providing a more comprehensive comparison of performance. Experimental results demonstrate that Sparse ICA outperforms these algorithms in terms of both accuracy and robustness. Furthermore, we applied Sparse ICA to real non-human primate (NHP) fMRI data, where it successfully identified sparse and independent functional networks in the marmoset brain, highlighting its effectiveness in analyzing complex fMRI data.

\section{METHODOLOGY}
\label{sec:med}
\subsection{Dataset}
\noindent\textbf{Synthetic Data:}  We constructed three distinct ground-truth source signals, each represented as a \(33 \times 33\) image and extended over 50 time frames (\(t = 1,\dots,50\)). These sources reflect three independent spatial patterns, shaped to resemble the digits ``1'', ``2 2'', and ``3 3 3''. The pixel values within these digit-like regions ranged between 0.5 and 1, with all remaining pixels set to zero.

To simulate realistic noise, structured Gaussian noise was added in two phases, following~\cite{wang2024sparse,risk2021simultaneous}. At \(t = 1\), a smoothed Gaussian random field (SD = 1, FWHM = 6) was generated. For \(t = 2\) to 50, noise followed an AR(1) process by scaling the previous frame by 0.47 and adding new smoothed noise (FWHM = 6). The signal-to-noise ratio (SNR) was controlled by adjusting the noise variance \(\sigma^2\), using \(\text{SNR} = \frac{1}{T \cdot \sigma^2} \sum_{i=1}^{Q} \lambda_i\), where \(\lambda_i\) are the nonzero eigenvalues of the source covariance. We set the SNR to 0.4 for all simulations.

\noindent\textbf{NHP Neuroimaging:} We utilized the marmoset fMRI datasets~\cite{Tian12nc} from the National Institutes of Health (NIH), USA, and the Institute of Neuroscience (ION), China, which together form a large-scale awake resting-state fMRI resource for the common marmoset (Callithrix jacchus). This study involved 16 animals, with 10 obtained from the NIH and 6 from the ION, with age ranges of approximately 4$\pm$2 years (NIH) and 3$\pm$1 years (ION). Across both sites, a total of 710 high-quality resting-state fMRI runs were collected, resulting in over 11,000 minutes of scan time. fMRI data were acquired using 7T (NIH) and 9.4T (ION) systems with custom-built coils, with a spatial resolution of 0.5 mm isotropic and a repetition time (TR) of 2 seconds. Standard preprocessing, including motion correction, spatial normalization to the Marmoset Brain Mapping (MBM V3) atlas~\cite{Liu21NI}, and spatial smoothing, was applied prior to Sparse ICA analysis.

\begin{figure}[!ht]
    \centering
    \includegraphics[width=0.5\textwidth, height=10cm]{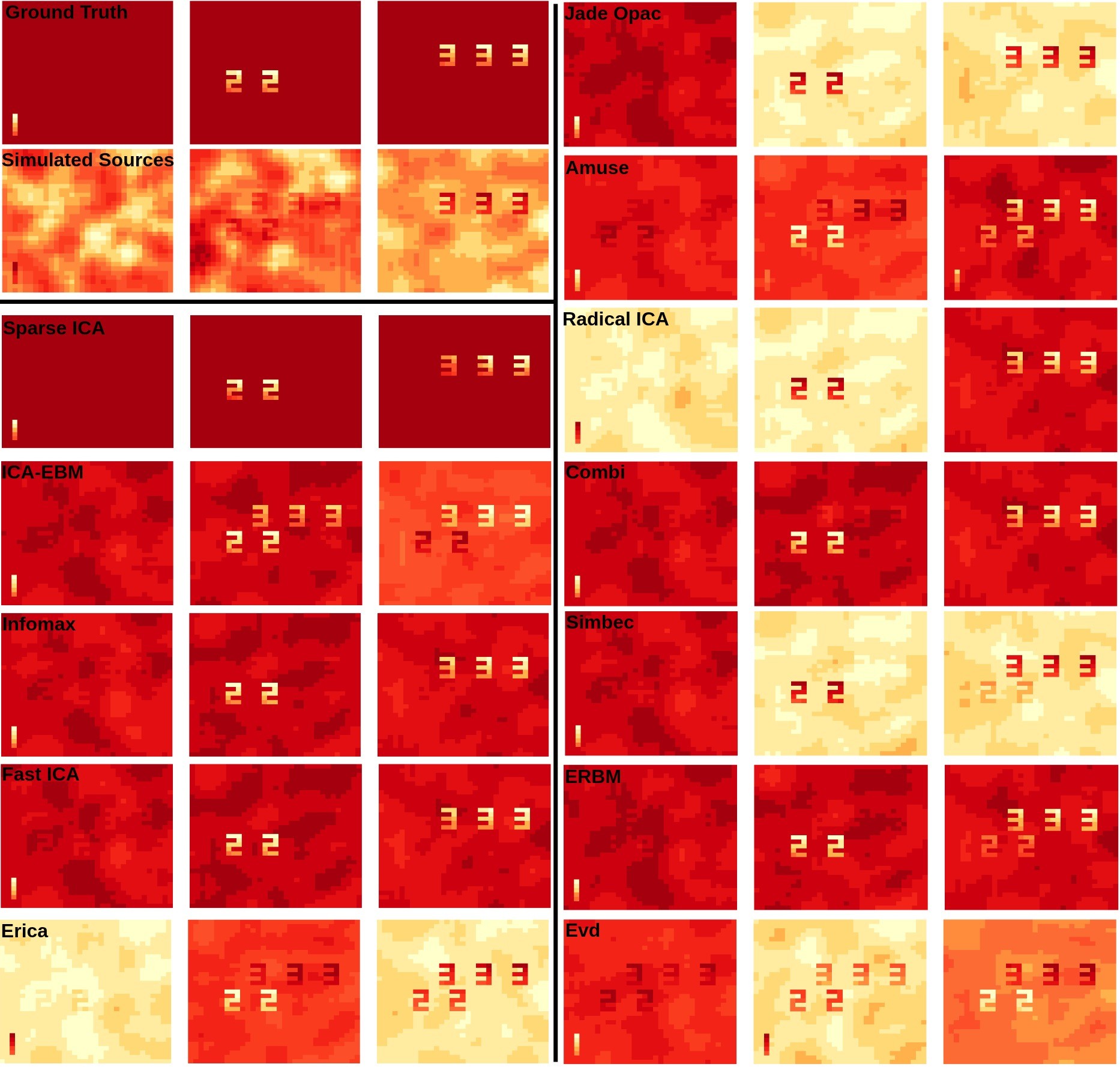}
    \caption{\textbf{Comparison of Sparse ICA with 11 other ICA algorithms.} Simulated ground truth sources and their mixtures are shown for reference. To evaluate Sparse ICA’s performance, we compared its results with those of 11 established ICA algorithms: Infomax, Fast ICA, Erica, Simbec, Evd, Jade Opac, Amuse, Radical ICA, Combi, ICA-EBM, and ERBM. The results demonstrate sparse ICA’s superior accuracy and robustness in recovering independent sources compared to these methods.}
    \label{fig:1}
\end{figure}

\begin{figure*}[!ht]
    \centering
    \includegraphics[width=\textwidth, height=4.5cm]{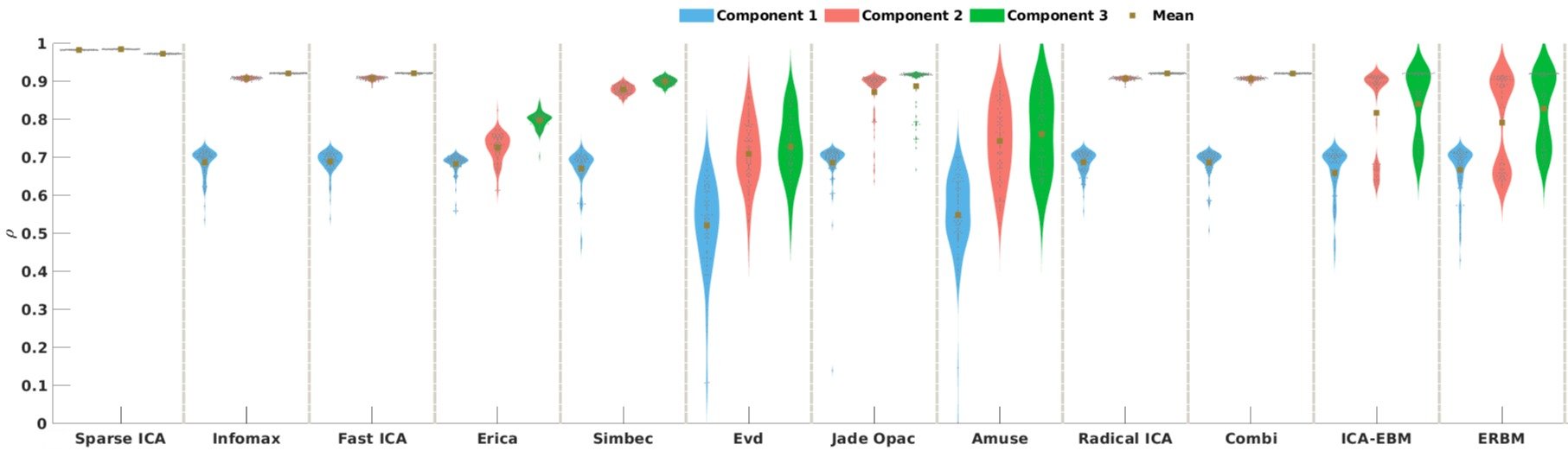}
    \caption{\textbf{Quantitative evaluation of Sparse ICA compared to 11 other ICA algorithms.} To assess the performance of Sparse ICA, we conducted 90 repeated simulation trials. In each trial, spatial correlation was measured between the estimated components and the ground truth across three source signals. The average results for Sparse ICA and 11 other ICA algorithms (Infomax, fast ICA, Erica, Simbec, Evd, Jade Opac, Amuse, radical ICA, Combi, ICA-EBM, and ERBM) are presented. This evaluation highlights the consistency and accuracy of Sparse ICA in recovering independent sources across repeated experiments.}
    \label{fig:2}
\end{figure*}

\begin{table*}[!ht]
\centering
\caption{\textbf{Benchmarking Sparse ICA against 11 widely used ICA algorithms.} Here presents the average spatial similarity (along with standard deviation) between the estimated and ground truth components across 90 repeated trials for Sparse ICA and 11 other ICA algorithms. An ICA algorithm is considered deterministic if it consistently produces the exact same output when given identical input data and parameters. In contrast, a nondeterministic algorithm may yield varying results across runs due to internal sources of randomness.}
\scalebox{0.89}{
\begin{tabular}{@{}llllc@{}}
\toprule
Class                                                                                                                                                                                                             & Algorithms                       & References                                                                                                    & Types                                  & Spatial Similarities  \\ \midrule
\multicolumn{1}{|l|}{\multirow{4}{*}{\begin{tabular}[c]{@{}l@{}}Maximum  Likelihood Based \\ (Estimate at most one Gaussian)\end{tabular}}}                                                                        & \multicolumn{1}{l|}{Infomax}     & \multicolumn{1}{l|}{\begin{tabular}[c]{@{}l@{}}Bell et al., 1995~\cite{bell1995information}\\ Correa et al., 2007~\cite{Correa2007PerformanceOB}\end{tabular}}         & \multicolumn{1}{l|}{non deterministic} & \multicolumn{1}{c|}{$0.84\pm0.01$} \\ \cmidrule(l){2-5} 
\multicolumn{1}{|l|}{}                                                                                                                                                                                            & \multicolumn{1}{l|}{Fast ICA}     & \multicolumn{1}{l|}{Hyvärinen et al., 1999~\cite{hyvarinen1999fast}}                                                                   & \multicolumn{1}{l|}{non deterministic} & \multicolumn{1}{c|}{$0.84\pm0.02$} \\ \cmidrule(l){2-5} 
\multicolumn{1}{|l|}{}                                                                                                                                                                                            & \multicolumn{1}{l|}{Radical ICA} & \multicolumn{1}{l|}{\begin{tabular}[c]{@{}l@{}}Learned-Miller et al., 2003~\cite{LearnedMiller03} \end{tabular}} & \multicolumn{1}{l|}{deterministic}     & \multicolumn{1}{c|}{$0.74\pm0.03$} \\ \cmidrule(l){2-5} 
\multicolumn{1}{|l|}{}                                                                                                                                                                                            & \multicolumn{1}{l|}{ICA-EBM}     & \multicolumn{1}{l|}{\begin{tabular}[c]{@{}l@{}}Li et al., 2010~\cite{XiLin10} \end{tabular}}               & \multicolumn{1}{l|}{non deterministic} & \multicolumn{1}{c|}{$0.82\pm0.02$} \\ \midrule
\multicolumn{1}{|l|}{\multirow{4}{*}{\begin{tabular}[c]{@{}l@{}}Maximum Likelihood Based\\ (Separates Gaussians with\\ different sample dependence \\ structure, i.e., autocorrelation\\ matrices)\end{tabular}}} & \multicolumn{1}{l|}{Evd}         & \multicolumn{1}{l|}{Georgiev et al., 2001~\cite{georgiev2001blind}}                                                                    & \multicolumn{1}{l|}{non deterministic} & \multicolumn{1}{c|}{$0.65\pm 0.10$} \\ \cmidrule(l){2-5} 
\multicolumn{1}{|l|}{}                                                                                                                                                                                            & \multicolumn{1}{l|}{ERBM}        & \multicolumn{1}{l|}{\begin{tabular}[c]{@{}l@{}}Fu et al., 2015~\cite{Fu2015sp}\end{tabular}}       & \multicolumn{1}{l|}{non deterministic} & \multicolumn{1}{c|}{$0.82\pm0.06$} \\ \cmidrule(l){2-5} 
\multicolumn{1}{|l|}{}                                                                                                                                                                                            & \multicolumn{1}{l|}{Amuse}       & \multicolumn{1}{l|}{\begin{tabular}[c]{@{}l@{}}Tong et al., 1990, 1991~\cite{tong1990amuse,tong1991amuse}\end{tabular}}  & \multicolumn{1}{l|}{deterministic}     & \multicolumn{1}{c|}{$0.69\pm0.10$} \\ \cmidrule(l){2-5} 
\multicolumn{1}{|l|}{}                                                                                                                                                                                            & \multicolumn{1}{l|}{Combi}       & \multicolumn{1}{l|}{\begin{tabular}[c]{@{}l@{}}Tichavsky et al., 2006, 2011~\cite{Tichavsky06,Tichavsky11}\end{tabular}}      & \multicolumn{1}{l|}{non deterministic} & \multicolumn{1}{c|}{$0.84\pm0.01$} \\ \midrule
\multicolumn{1}{|l|}{\multirow{3}{*}{\begin{tabular}[c]{@{}l@{}}Cumulant-based (Estimate at most\\ one Gaussian)\end{tabular}}}                                                                                   & \multicolumn{1}{l|}{Simbec}      & \multicolumn{1}{l|}{Cruces et al., 2001~\cite{Cruces01}}                                                                      & \multicolumn{1}{l|}{deterministic}     & \multicolumn{1}{c|}{$0.84\pm0.01$} \\ \cmidrule(l){2-5} 
\multicolumn{1}{|l|}{}                                                                                                                                                                                            & \multicolumn{1}{l|}{Erica}       & \multicolumn{1}{l|}{Cruces et al., 2002~\cite{CRUCES200287}}                                                                      & \multicolumn{1}{l|}{deterministic}     & \multicolumn{1}{c|}{$0.77\pm0.09$} \\ \cmidrule(l){2-5} 
\multicolumn{1}{|l|}{}                                                                                                                                                                                            & \multicolumn{1}{l|}{Jade Opac}   & \multicolumn{1}{l|}{\begin{tabular}[c]{@{}l@{}}Cardoso et al., 1993~\cite{cardoso1993blind}\end{tabular}}        & \multicolumn{1}{l|}{deterministic}     & \multicolumn{1}{c|}{$0.76\pm0.09$} \\ \midrule
Relax-and-Split Sparse ICA      & Sparse ICA   & Wang et al., 2024~\cite{wang2024sparse} & non deterministic & \multicolumn{1}{c}{$0.96\pm0.02$}  \\ \bottomrule
\end{tabular}}
\label{tab}
\end{table*}

\subsection{Relax-and-Split Sparse ICA (Sparse ICA)}
Assuming the input is a whitened data matrix \( \tilde{\mathbf{X}} \in \mathbb{R}^{P \times Q} \), where \( P \) is the number of spatial locations and \( Q \) is the number of components. The goal is to decompose \( \tilde{\mathbf{X}} \) into a sparse spatial map matrix \( \mathbf{V} \in \mathbb{R}^{P \times Q} \) and an orthogonal rotation matrix \( \mathbf{U} \in \mathbb{R}^{Q \times Q} \), such that \( \tilde{\mathbf{X}} \approx \mathbf{V} \mathbf{U}^\top \). Sparsity is enforced on \( \mathbf{V} \) using an \( \ell_1 \) penalty, and \( \mathbf{U} \) is constrained to lie on the Stiefel manifold \( \mathcal{O}_{Q \times Q} \), the set of orthogonal matrices.

The algorithm solves the following optimization problem:
\begin{equation}
\min_{\mathbf{V},\ \mathbf{U} \in \mathcal{O}_{Q \times Q}} \frac{1}{2} \left\| \tilde{\mathbf{X}} - \mathbf{V} \mathbf{U}^\top \right\|_F^2 + \nu \left\| \mathbf{V} \right\|_1
\end{equation}

\noindent The optimization is carried out via alternating minimization:

\begin{itemize}
    \item \textbf{(V-step) Soft-thresholding update:} Each element of \( \mathbf{V} \) is updated using a soft-thresholding rule applied to \( \tilde{\mathbf{X}} \mathbf{U} \):
    \begin{equation}
    V_{ij} \leftarrow \left( \left| [\tilde{\mathbf{X}} \mathbf{U}]_{ij} \right| - \sqrt{2\nu} \right)_+ \cdot \text{sign}([\tilde{\mathbf{X}} \mathbf{U}]_{ij})
    \end{equation}

    \item \textbf{(U-step) Orthogonal Procrustes update:} Update \( \mathbf{U} \) via SVD of the product \( \tilde{\mathbf{X}}^\top \mathbf{V} \):
    \begin{equation}
    \tilde{\mathbf{X}}^\top \mathbf{V} = \tilde{\mathbf{U}} \tilde{\Sigma} \tilde{\mathbf{V}}^\top, \quad \mathbf{U} \leftarrow \tilde{\mathbf{U}} \tilde{\mathbf{V}}^\top
    \end{equation}
\end{itemize}
Since both updates have closed-form solutions, the objective function values decrease monotonically across iterations, ensuring convergence. The detailed proof can be found in prior work~\cite{wang2024sparse}.

\section{RESULTS}
\label{sec:res}
\begin{figure*}[!ht]
    \centering
    \includegraphics[width=0.9\textwidth, height=12.5cm]{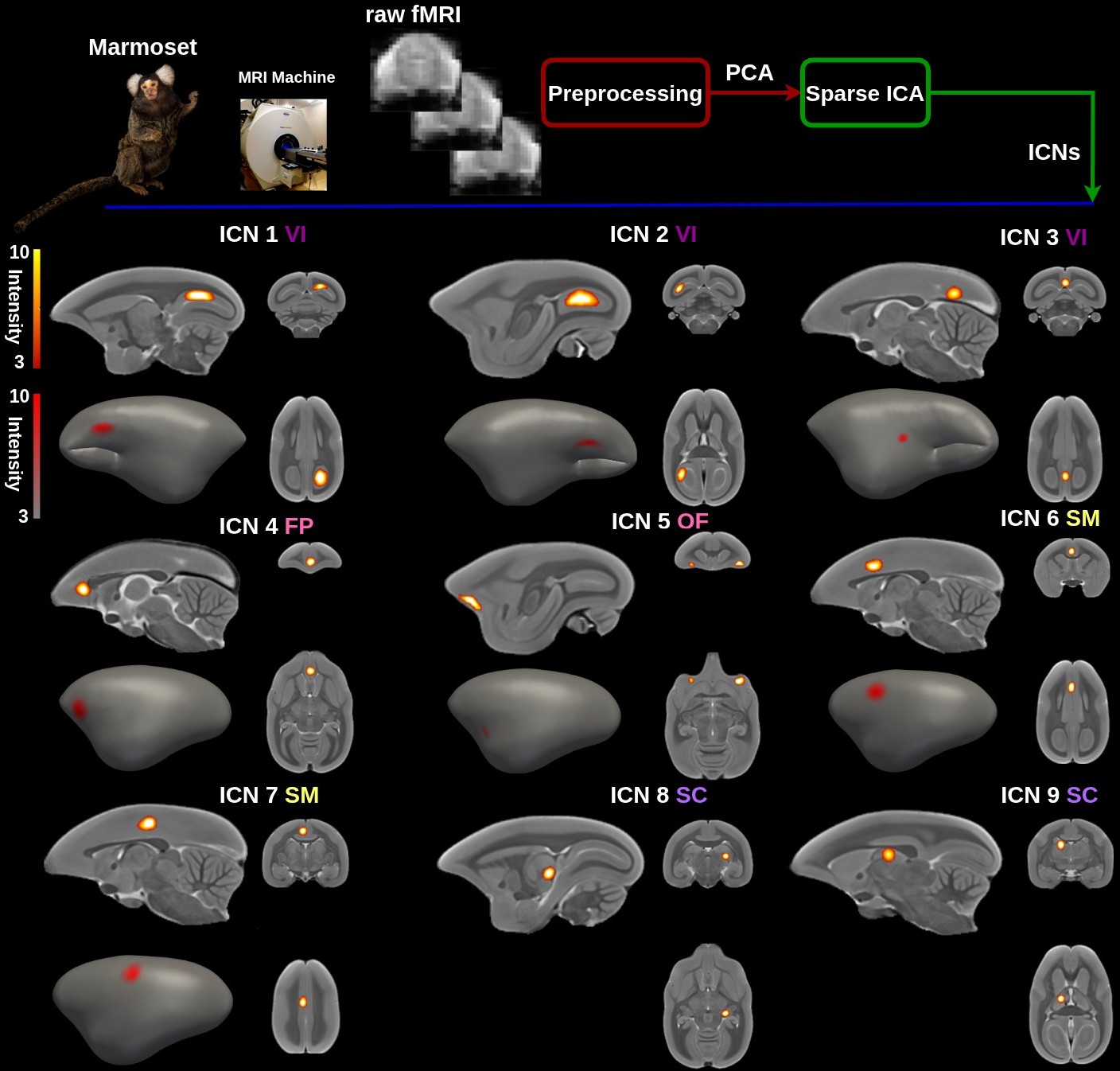}
    \caption{\textbf{Decomposition of independent component networks from marmoset resting-state brain using Sparse ICA.} Sparse ICA was applied to marmoset fMRI data, resulting in 9 selected independent components representing distinct brain networks. These estimated components are both sparse and statistically independent, enhancing interpretability and specificity of the functional networks identified.}
    \label{fig:3}
\end{figure*}

In Fig.\ref{fig:1}, the unmixing performance of Sparse ICA is evaluated against 11 other ICA algorithms (listed in Table.\ref{tab}) in separating mixed source signals, with the ground truth components included for comparison. Overall, Sparse ICA demonstrates superior performance across all three components relative to the other methods. Among the remaining ICA approaches, Infomax and fast ICA still show relatively strong results, outperforming many of the others. In contrast, methods such as Simbec, Amuse, and Evd exhibit noticeably poorer separation, often failing to accurately reconstruct the shape or structural features of the original sources.

To complement the qualitative evaluation, we performed 90 independent simulation runs for each ICA method to quantitatively assess their consistency. The outcomes, illustrated in Fig.\ref{fig:2}, support the visual findings, Sparse ICA consistently delivers better performance across all repetitions. This underscores the potential advantage of approaches that leverage shared source structure in enhancing unmixing accuracy. Furthermore, Table.\ref{tab} provides a summary of the mean and standard deviation of spatial correlation scores for each algorithm, reinforcing the robustness and reliability of Sparse ICA in recovering the underlying components.

In Fig.\ref{fig:3}, nine selected resting-state independent component networks (ICNs) from the marmoset brain are shown using both volumetric and inflated surface representations. These ICNs were derived using sparse ICA, which promotes spatial sparsity during the decomposition process, resulting in more localized and interpretable network structures.

The identified components include the Visual Network (VI; ICN1-ICN3), the Frontal Pole (FP; ICN4), the Orbitofrontal Network (OF; ICN5), the Sensorimotor Network (SM; ICN6-ICN7), and the Subcortical Network (SC; ICN8-ICN9). Each of these networks demonstrates a clear and focused spatial distribution, consistent with known functional brain networks in the marmoset. The ability of Sparse ICA to recover such distinct and non-overlapping ICNs highlights its effectiveness in isolating meaningful functional structures, especially in non-human primate neuroimaging studies.

\section{ACKNOWLEDGMENTS}
The authors declare that there are no conflicts of interest related to this research. This work was supported by NSF grant 2112455, and NIH grants R01MH123610 and R01MH119251. We also thank Zihang Wang from Emory University for his comments on Sparse ICA.

\clearpage
\section{COMPLIANCE WITH ETHICAL STANDARDS}
This research study was conducted retrospectively using non-human primate data obtained through open-access sources. Ethical approval was not required, as confirmed by the license accompanying the open access data.

\bibliographystyle{IEEEbib}
\bibliography{refs}

\end{document}